\documentclass[aps,preprint,showpacs]{revtex4}
\usepackage{amsfonts}
\usepackage{amsmath}
\usepackage{amssymb}
\usepackage{graphicx}

\begin{document}
\preprint{MYA/00133}
\title{Calculation of the photoionization with de-excitation cross sections
of He and helium-like ions}
\author{M. Ya. Amusia}
\affiliation{Racah Institute of Physics, The Hebrew University,
Jerusalem 91904, Israel; A. F. Ioffe Physical-Technical Institute,
St. Petersburg, 194021, Russia}
\author{R. Krivec}
\affiliation{Department of Theoretical Physics, J. Stefan Institute,
P.O. Box 3000, 1001 Ljubljana, Slovenia}
\author{E. Z. Liverts}
\affiliation{Racah Institute of Physics, The Hebrew University,
Jerusalem 91904, Israel}
\author{V. B. Mandelzweig}
\affiliation{Racah Institute of Physics, The Hebrew University,
Jerusalem 91904, Israel}
\pacs{32.80.Fb, 31.15.Ja}

\begin{abstract}
We discuss the results of the calculation of the photoionization with
de-excitation of excited He and helium-like ions Li$^{+}$ and B$^{3+}$ at
high but non-relativistic photon energies $\omega $. Several lower $^{1}S$
and $^{3}S$ states are considered. We present and analyze the ratios
$R_{d}^{+\ast}$ of the cross sections of photoionization with
de-excitation, $\sigma_{(d)}^{+\ast}(\omega)$, and of the photo-ionization
with excitation, $\sigma ^{+\ast}(\omega )$. The dependence of
$R_{d}^{+\ast}$ on the excitation of the target object and the charge of its
nucleus is presented. Apart to theoretical interest, results obtained can be
verified using such long living excited state as $2^{3}S$ of He.
\end{abstract}

\maketitle
\begin{table}[hbt]
\begin{center}
\caption{Values of the ratios $I_{n_{f}n_{i}} $ and  $R^{+*}_{id}$
for the singlet states of the helium atom.}
\begin{tabular}{|c|c|c|c|c|c|c|c|}
\hline \( n_{f}\, \setminus \, n_{i} \)& 1& 2& 3& 4& 5& 6&7\\
\hline \hline 1& \textbf{0.9295}& 0.0493& 0.0136&
0.0055&0.0028&0.0016&
0.0010\\
\hline 2& 0.0446& \textbf{.5346}& 0.0702& 0.0231& 0.0106& 0.0059&
0.0036\\
\hline 3& 0.0055& 0.3993& \textbf{.1668}& 0.0533& 0.0237& 0.0131&
0.0079\\
\hline 4& 0.0018& 0.0035& 0.7319& \textbf{.0019}& 0.0063& 0.0051&
0.0037\\
\hline 5& 0.0008& 0.0017& 0.0131& 0.7761& \textbf{.0594}& 0.0107&
0.0029\\
\hline 6& 0.0005& 0.0009& 0.0001& 0.1380& 0.4997& \textbf{.1430}&
0.0541\\
\hline 7& 0.0003& 0.0005& 0.0001& 0.0003& 0.3865& 0.1606&\textbf{.1129}\\
\hline 8& 0.0002& 0.0003& 0.0001& & 0.0103& 0.5996&0.0025\\
\hline 9& 0.0001& 0.0002& 0.0001& & & 0.0597&0.6106\\
\hline 10& 0.0001& 0.0002& 0.0001& & & 0.0001&0.1969\\
\hline 11& 0.0001& 0.0001& & & & &0.0035\\
\hline 12& & 0.0001& & & & &0.0000\\
\hline\hline{$R^{+*}_{id}$}&&\textbf{.0523}& \textbf{.0917}&
\textbf{.0893}&\textbf{.0452}&\textbf{.0378}&\textbf{.0790}\\ \hline
\end{tabular}
\end{center}
\end{table}

\begin{table}[hbt]
\begin{center}
\caption{The same as in Table I, but for the triplet states of the
helium atom.}
\begin{tabular}{|c|c|c|c|c|c|c|}\hline
\( n_{f}\, \setminus \, n_{i} \)& 2& 3& 4& 5& 6&7\\
\hline \hline 1& .0338& .0087& .0034& .0017& .0009&.0006\\\hline
2&\textbf{.7823}& .0591& .0170& .0073& .0039&.0023\\\hline
3&.1733&\textbf{.4053}& .0902& .0352& .0177&.0103\\\hline
4&.0044&.5231& \textbf{.0760}& .0419& .0228&.0137\\\hline
5&.0014&.0006&.7595&\textbf{.0059}& .0005&.0016\\\hline
6&.0006&.0004&.0526&.6723&\textbf{0.0995}&.0247\\\hline
7&.0003&.0003& .0001& .2324& .3434&\textbf{.1470}\\\hline
8&.0002&.0002& &.0027&.4812&.0615\\\hline
9&.0001&.0001&&&.0299&.6144\\\hline 10&.0001&.0001& &&&.1225\\\hline
11& .0001& .0001& & & &.0012\\\hline 12& & & & & &.0000\\
\hline \hline $R^{+*}_{id}$& \textbf{.0350}& \textbf{.0728}&
\textbf{.124}& \textbf{.0942}& \textbf{.0479}&\textbf{.0562}\\
\hline
\end{tabular}
\end{center}
\end{table}

\begin{table}[hbt]
\begin{center}
\caption{Values of the ratios $I_{n_{f}n_{i}}$ and $R^{+*}_{id}$ for
the singlet states of the Li$^{+} $ ion.}
\begin{tabular}{|c|c|c|c|c|c|c|c|} \hline
\(n_{f}\, \setminus \,n_{i}\)& 1& 2& 3& 4& 5& 6&7\\\hline \hline
1&\textbf{.9716}&.0718&.0210&.0087&.0044&.0025&.0016\\\hline
2&.0157&\textbf{.7456}&.0516&.0147&.0064&.0034&.0020\\\hline
3&.0023&.1528& \textbf{.5476}&.0835&.0293&.0141&.0080\\\hline
4&.0008&.0102&.3561& \textbf{.3000}&.0813&.0347&.0183\\\hline
5&.0004&.0032&.0082&.5824&\textbf{.1053}&.0523&.0276\\\hline
6&.0002&.0015&.0029&.0009&.7627&\textbf{.0100}&.0177\\\hline
7&.0001&.0008&.0014&.0013&.0052&.8381&\textbf{.0107}\\\hline
8&.0001&.0005&.0008&.0008&.0003&.0417&.7807\\\hline
9&.0001&.0003&.0005&.0005&.0003&.0000&.1314\\\hline
10&.0000&.0002&.0003&.0003&.0002&.0001&.0000\\\hline
11&.0000&.0002&.0002&.0002&.0002&.0001&.0000\\\hline
12&.0000&.0001&.0002&.0002&.0001&.0001&.0000\\\hline \hline {\(
R^{+*}_{id} \)}& & \textbf{.0784}& \textbf{.0790}& \textbf{.120}&
\textbf{.139}& \textbf{.120}&\textbf{.0815}\\
\hline
\end{tabular}
\end{center}
\end{table}

\begin{table}[hbt]
\begin{center}
\caption{The same as in Table III, but for the triplet states of
the Li$^{+} $ ion.}
\begin{tabular}{|c|c|c|c|c|c|c|}\hline
\(n_{f}\, \setminus \,n_{i}\) & 2 & 3 & 4 & 5 & 6 & 7 \\\hline
\hline 1 & .0559 & .0158 & .0064 & .0032 & .0018 & .0011
\\ \hline
2 & \textbf{.8762} & .0316 & .0079 & .0032 & .0017 & .0010 \\
\hline
3 & .0574 & \textbf{.7283} & .0771 & .0241 & .0109 & .0060 \\
\hline
4 & .0044 & .2085 & \textbf{.4788} & .0982 & .0375 & .0188 \\
\hline
5 & .0013 & .0070 & .4194 & \textbf{.2340} & .0836 & .0391 \\
\hline
6 & .0006 & .0022 & .0031 & .6321 & \textbf{.0649} & .0467 \\
\hline
7 & .0003 & .0010 & .0016 & .0003 & .7789 & \textbf{.0011} \\
\hline 8 & .0002 & .0005 & .0008 & .0006 & .0179 & .8094 \\
\hline 9 & .0001 & .0003 & .0005 & .0004 & .0001 & .0747 \\
\hline 10 & .0001 & .0002 & .0003 & .0003 & .0001 & .0000 \\
\hline 11 & .0001 & .0002 & .0002 & .0002 & .0002 & .0001 \\
\hline 12 & .0000 & .0001 & .0001 & .0001 & .0001 & .0000 \\
\hline\hline {$R^{+*}_{id} $} & \textbf{.0594} & \textbf{.0499} &
\textbf{.101} & \textbf{.148} & \textbf{.157} & \textbf{.127} \\
\hline
\end{tabular}
\end{center}
\end{table}

\begin{table}[hbt]
\begin{center}
\caption{Values of the ratios $I_{n_{f}n_{i}}$ and $R^{+*}_{id}$ for
the singlet states of the B$^{3+}$ ion.}
\begin{tabular}{|c|c|c|c|c|c|c|c|}\hline
\(n_{f}\, \setminus \,n_{i}\)& 1 & 2 & 3 & 4 & 5 & 6 &7
\\ \hline\hline
1&\textbf{.9905}&.0892&.0270&.0114&.0058&.0034&.0020
\\ \hline
2&.0047&\textbf{.8485}&.0235&.0057&.0023&.0012&.0007
\\ \hline
3&.0008&.0448&\textbf{.8156}&.0495&.0138&.0060&.0031
\\ \hline
4&.0003&.0054&.1098&\textbf{.7053}&.0725&.0228&.0102
\\ \hline
5&.0001&.0018&.0094&.2009&\textbf{.5663}&.0877&.0300
\\ \hline
6&.0001&.0009&.0030&.0116&.3131&\textbf{.4197}&.0900
\\ \hline
7&.0000&.0005&.0014&.0036&.0112&.4376&\textbf{.2746}
\\ \hline
8&.0000&.0003&.0008&.0017&.0037&.0080&.5493
\\ \hline
9&.0000&.0002&.0005&.0010&.0018&.0030&.0036
\\ \hline
10&.0000&.0001&.0003&.0006&.0009&.0019&.0015
\\ \hline
11&.0000&.0001&.0002&.0004&.0007&.0006&.0036
\\ \hline
12&.0000&.0001&.0002&.0003&.0004&.0007&.0000
\\ \hline\hline
{$R^{+*}_{id}$}&&\textbf{.0987}&\textbf{.0536}&
\textbf{.0718}&\textbf{.105}&\textbf{.138}&\textbf{.158}
\\ \hline
\end{tabular}
\end{center}
\end{table}

\begin{table}[hbt]
\begin{center}
\caption{The same as in Table V, but for the triplet states of
the B$^{3+} $ ion.}
\begin{tabular}{|c|c|c|c|c|c|c|}
\hline
\(n_{f}\, \setminus \,n_{i}\) & 2 & 3 & 4 & 5 & 6 & 7 \\
\hline\hline 1 & .0764 & .0228 & .0095 & .0048 & .0028 & .0017
\\ \hline
2 & \textbf{.9025} & .0122 & .0027 & .0010 & .0005 & .0003 \\
\hline
3 & 0.0162 & \textbf{.8914} & .0359 & .0091 & .0038 & .0020 \\
\hline
4 & .0019 & .0613 & \textbf{.8004} & .0622 & .0179 & .0078 \\
\hline
5 & .0006 & .0054 & .1337 & \textbf{.6722} & .0840 & .0272 \\
\hline
6 & .0003 & .0017 & .0084 & .2306 & \textbf{.5265} & .0964 \\
\hline
7 & .0002 & .0008 & .0025 & .0095 & .3460 & \textbf{.3783} \\
\hline 8 & .0001 & .0004 & .0012 & .0030 & .0082 & .4717 \\
\hline 9 & .0001 & .0003 & .0006 & .0014 & .0028 & .0045 \\
\hline 10 & .0000 & .0002 & .0004 & .0008 & .0014 & .0023 \\
\hline 11 & .0000 & .0001 & .0003 & .0005 & .0008 & .0013 \\
\hline 12 & .0000 & .0001 & .0002 & .0003 & .0005 & .0005 \\
\hline\hline {$R^{+*}_{id} $} & \textbf{.0829} & \textbf{.0364} &
\textbf{.0507} & \textbf{.0839} & \textbf{.123} & \textbf{.157}
\\ \hline
\end{tabular}
\end{center}
\end{table}
\section{Introduction}

The processes of two-electron photoionization and ionization with excitation
have attracted the attention of theorists and experimentalists for a long
time. A steady increase of activity has occurred during the last decade
\cite{1,2,3,4,5,6,7,8}. The interest in these processes is motivated to a
large extent by the desire to test our ability to calculate reliably the
two-electron wave function and to understand the mechanisms of these
processes that take place solely due to the interelectron interaction.

The simplest objects where the interelectron interaction can manifest itself
are the two-electron systems as the He atom and the helium-like ions.
Therefore they represent the main targets of investigation. Recently a
number of studies have been carried out of the two-electron photoionization
cross sections of these systems, including the corresponding ratios of the
two-electron and the single-electron cross sections, at high but
nonrelativistic photon energies $\omega $ (see \cite{9,10,11,12} and
references therein). 

In high photon energy region the cross sections of two-electron processes can be
expressed via the initial state wave functions. The initial states
considered were the ground and excited states of He and the helium-like
ions. 

For high but nonrelativistic $\omega $, the dominating mechanisms of
the two-electron ionization and ionization with excitation are twofold:
shake-off and the initial state correlations. Both exhibit the same
dependence on $\omega$ at high $\omega$. The contribution of the final-state
interaction, where the second electron is excited or ionized due to a
collision with the primary eliminated electron that absorbs the incoming
photon, decreases faster with $\omega$ than shake-off. The quasi-free
mechanism \cite{13} operates in the situation where both electrons are
ionized and acquire almost equal energies. Therefore the quasi-free
mechanism is not taken into consideration in the framework of the ionization
with excitation or de-excitation.

If the initial state is excited, the elimination of one of the electrons can
be accompanied not only by the excitation of the second one, but also by
de-excitation. As far as we are aware, the de-excitation process almost
completely escaped theoretical investigation. However, it is expressed by integrals
similar  but not identical to those entering the expressions of the two-electron ionization and
ionization with excitation cross sections \cite{14,9,10,11}. The aim of this paper is to
study the photoionization accompanied by de-excitation. In principle, this
process can be separated experimentally from the other two-electron
processes, i.e., the double ionization and ionization with excitation, if
the photoelectron's energy for the given incoming photon frequency $\omega$
is detected.

Here we obtain the initial state wave function using the correlation function hyperspherical harmonic method
(CFHHM). The local accuracy of this wave function was previously studied in Ref.[15] for the ground and
the $2^{1}S$ state of the helium atom, where it was shown that the local
deviation of CFHHM wave function from the exact value is extremely small.
Very accurate nonvariational CFHHM wave functions of the He atom and the
helium-like ions in their ground and several lowest excited $^{1}S$ and
$^{3}S$ states \cite{16,17} were utilized therefore to calculate the cross
sections of the processes of interest. We calculated the high-energy
photoionization cross sections that can be expressed solely via the initial state
two-electron wave functions $\Psi_{i}(\mathbf{r}_{1},\mathbf{r}_{2})$. 

In this work we will use these nonvariational wave functions for the He atom and
the helium-like ions in several lowest excited $^{1}S$ and $^{3}S$ states to
calculate the high-energy limits of the cross sections of photoionization
with de-excitation. The results of these limits will be compared with the cross sections of
the photoionization with excitation and with the single-electron
photoionization cross sections. To study the theoretically interesting nuclear 
charge dependences of considered values, we investigate also the Li$^{+}$ and B$^{3+}$ ions.

Unfortunately, till now in absolute majority of experiments only the double- and
single-charged ions are counted. Therefore, the excitation and de-excitation 
processes remained not detected but included into the yield of single charge ions.

\section{Main formulas}

We start from the formula for the two-electron photoionization cross section
at asymptotically high $\omega$ obtained in \cite{9} that has some recent derivations
\cite{10,11}. The expression for the cross section $\sigma^{+*}(\omega)$ of
the ionization with excitation, in this $\omega$ region, for the He atom and
the helium-like ions in their excited states, is accordingly as follows:
\begin{equation} \sigma_{i}^{+*}(\omega
)=\frac{32Z^{2}\sqrt{2}\pi^{2}}{3c\omega ^{7/2}} \sum
_{n_{f}}I_{n_{f}n_{i}},
\end{equation}
where $I_{n_{f}n_{i}}$ is the overlap integral defined as
\begin{equation}
{I_{n_{f}n_{i}}= 4\pi \mid \int ^{\infty }_{_{0}}\Psi
_{i}(0,r)R_{n_{f}0}(r)r^{2}dr\mid ^{2} }.
\end{equation}

Here $Z$ is the nuclear charge, $i(n_{i})$ denotes the initial state, 
$R_{n_{f}0}(r)$ are the hydrogenic single-electron radial wave
functions with the principal quantum number $n_{f}$ and angular momentum
zero. We note that the excitations of states with nonzero angular momenta $l$
decrease faster than $\omega^{-7/2}$: namely as $\omega^{(-7/2+l)}$ .

The photoionization cross section $\sigma^{+}(\omega)$ of the inner
electron, without alteration of the state of the outer one, is given by the
following expression
\begin{equation}
\sigma_{i}^{+}(\omega )=\frac{128Z^{2}\sqrt{2}\pi ^{3}}{3c\omega ^{7/2}}
{\mid \int ^{\infty }_{_{0}}\Psi _{i}(0,r)R_{n_{i}0}(r)r^{2}dr\mid
^{2}}.
\end{equation}

The cross section $\sigma_{(d)}^{+\ast}$ of the photoionization with
de-excitation is calculated using the expressions (1) and (2), where
the summations over $n_{f}$ include different than for ionization with excitation values.
As was mentioned above, we classify the
initial state by its principal quantum number $n_{i}$. The ground state can
be considered approximately as a state with $n_{i}=1$. The next state is
$n_{i}=2$ and so on: the higher the excitation, the more precise becomes the
assignment of a given state to an integer principal quantum number. With
increasing excitation principal quantum number $n_{i}$, the wave function
approaches a symmetrized product of two pure Coulomb wave functions. One is
an $1${\large $s$} electron function in the nuclear field with charge $Z$, and the other is a function
in the $(Z-1)$ field. Thus $\sigma^{+\ast}(\omega)$ includes the summation
over $n_{f}>n_{i}$ while $\sigma _{i(d)}^{+\ast}(\omega)$ includes the
summation over $n_{f}<n_{i}$. The ratio $R_{id}^{+\ast}$ of the cross
section $\sigma_{i(d)}^{+\ast}(\omega)$ and the sum of
$\sigma_{i}^{+}(\omega)$ and $\sigma_{i}^{+\ast}(\omega)$ is given by the
expression:
\begin{equation}
R_{id}^{+\ast}{\equiv}
\frac{\sigma_{i(d)}^{+\ast}(\omega)}{\sigma_{i}^{+}(\omega)+\sigma_{i}^{+\ast}(\omega)}
=\frac{\sum_{n_{f}<n_{i}}I_{n_{f}n_{i}}}{\sum_{n_{f}\geq n_{i}}I_{n_{f}n_{i}}}.
\end{equation}

It is seen from Eq.(2) that $\sigma_{i}^{+\ast}$ and $\sigma_{i(d)}^{+\ast}$ 
probe different parts of initial state wave function $\Psi _{i}(0,r)$.

\section{Results of calculations}

The Tables present the results of our calculations for the singlet and
triplet states of the He atom (Tables I and II), and the helium-like ions
Li$^{+}$ (Tables II and IV) and B$^{3+}$ (Tables V and VI).

Most results can be obtained using a small hyperspherical harmonic (HH)
basis with $K_{m}=48 $, for some ground states even with $ K_{m}=40 $,
where $ K_{m} $ is the maximum global angular momentum used in the HH expansion.
The main problem is the calculation of the overlap integrals (2) between the accurate
CFHHM wave function and the hydrogenic single-electron ones of the highest orders.
The very precise calculations of the CFHHM wave function $ \Psi _{i}(0,r) $
at one electron-nucleus zero-distance for large values of $ r $ are required.
At that, the precision of the CFHHM wave function, including its values for
the large $ r $, increases slowly with increasing $ K_{m} $, while the
calculation time increases considerably with $ K_{m} $ growth. We used here
the HH basis with $ K_{m}=100 $ for the excited $ 2S-7S $ both singlet
and triplet states of He atom and the one with $ K_{m}=64 $ for the other
ions in the excited states, which make the calculation quite difficult.
This gives data with accuracy well above obtainable experimentally in foreseeable future.

In the Tables we presented only several the first terms $( n_{f}\leq 12) $
of the set of integrals (2). However, we have used 200 single-particle states
to calculate the cross-section ratios, so that the error of summation over states
at least is less than the error of the data presented in the Tables.

\section{Discussion and conclusions}

In all cases the probability of the initially excited electron to remain on
the same level ($n_{f}=n_{i}$) decreases with $n_{i}$. In principle, this is
quite natural since the higher the $n_{i}$, the easier it is to alter the
electron state in the course of the rapid elimination of the $1S$ electron.
This decrease is monotonic for pure hydrogenic functions; in the lowest
order in interelectron interaction, that result can be demonstrated
analytically \cite{18}. For the He atom the decrease is non-monotonic and
the overlap integral Eq.\ (2) reaches its minimum value at $n_{i}=n_{f}=4$
for the singlet states and at $n_{i}=n_{f}=5$ for the triplet states. This
is the consequence of the strong deviation of $\Psi_{i}({0},{r})$ from its
simplest approximation. A trace of the helium-like behavior can be found in
the minimum of $I_{n_{f}n_{i}}$ at $n_{i}=n_{f}=7$ for singlet states in
Li$^{+}$.

The B$^{3+}$ ion is already purely hydrogenlike, the role of its
interelectron interaction being relatively smaller than in He. This brings
about the decrease of $I_{n_{f}n_{i}}$ for $n_{f}=n_{i}$ with increasing
$n_{i}$. As a result, the probability for the initially excited electron to
remain on the same level after the inner $1s$ electron is photoionized
increases. Indeed, $I_{77}$ for B$^{3+}$ is about $30$ times bigger than for
Li$^{+}$. The corresponding ratio for the triplet $I_{77}$ values is even
larger by an additional factor of $15$. As seen from the Tables, the most
probable process is the ionization with excitation to the next one or two
levels. The ratio $R_{id}^{+\ast}$ varies relatively little, from $0.035$ to
$0.158$. It has quite a complex form, with at least two minima for the
He atom and at least one minimum for the singlet and triplet states of
B$^{3+}$.

Let us have in mind, that the process of photoionization with excitation can be relatively
easily distinguished from other two-electron processes, particularly from
the two-electron ionization. This can be achieved simply by detecting
photoelectrons with energies larger than the energy of the incoming photon.
The ratio $R_{id}^{+\ast}$ of the cross section of the ionization with
de-excitation and the cross section of the ionization with excitation
presents a rather complex function of the initial state and is completely
determined by its wave function. Observation of this process could serve as
an additional to studies of ionization with excitation, verification
of the quality of the wave functions employed in
describing the initial state of the target atoms or ions.

Note, that a target consisting of excited atoms can be produced by initial
illumination of a He gas volume by e.g. laser light. The $p$-excited states then
radiatively decay into excited $s$-states.

Targets of triplet excited states can be produced by colliding at small angles
of $\alpha$-particles beem with a magnetized, i.e. occupied by having electrons
with the same spin orientation metallic surface.

The development of experimental technique and growth of intensity of available high photon beams
will lead without doubt to the experiments, in which for a given photon frequency
the outgoing photoelectron's energy will be detected accurately enough.
It will permit to study ionization with excitation and de-excitation.
A good object for de-excitation studies would be the $2^{3}S$ state of helium
the lifetime of which is about eight minutes. Experimental detection of 
ionization with de-excitation is, of course, an experimental challenge 
and we do believe that it will attract experimentalists very soon.

\begin{acknowledgments}
We acknowledge the financial support of the Binational Science foundation
under the grant 2002064 and the Israeli Science foundation under the grant
174/03.
\end{acknowledgments}


\end{document}